\documentclass[preprint,showpacs,preprintnumbers,amsmath,amssymb,superscriptaddress]{revtex4}

\usepackage{graphicx}
\usepackage{dcolumn}
\usepackage{bm}
\usepackage{color}
\usepackage{ulem}

\def\prtx{PrT$_2$X$_{20}$}

\def\prruzn{PrRu$_2$Zn$_{20}$}
\def\prirzn{PrIr$_2$Zn$_{20}$}
\def\prrhzn{PrRh$_2$Zn$_{20}$}
\def\larhzn{LaRh$_2$Zn$_{20}$}
\def\lairzn{LaIr$_2$Zn$_{20}$}
\def\laruzn{LaRu$_2$Zn$_{20}$}

\def\prpb{PrPb$_{3}$}
\def\prossb{PrOs$_{4}$Sb$_{12}$}

\def\tq{$T_{\rm Q}$}

\def\tc{$T_{\rm c}$}

\def\ts{$T_{\rm s}$}

\begin{document}

\preprint{APS/123-QED}

\title{Simultaneous Superconducting and Antiferroquadrupolar Transitions in {\prrhzn}}

\author{T. Onimaru}
\email{onimaru@hiroshima-u.ac.jp}
\affiliation{%
Department of Quantum Matter, Graduate School of Advanced Sciences of Matter, Hiroshima University, Higashi-Hiroshima 739-8530, Japan
}%
\author{N. Nagasawa}%
\affiliation{%
Department of Quantum Matter, Graduate School of Advanced Sciences of Matter, Hiroshima University, Higashi-Hiroshima 739-8530, Japan
}%
\author{K. T. Matsumoto}%
\affiliation{%
Department of Quantum Matter, Graduate School of Advanced Sciences of Matter, Hiroshima University, Higashi-Hiroshima 739-8530, Japan
}%
\author{K. Wakiya}%
\affiliation{%
Department of Quantum Matter, Graduate School of Advanced Sciences of Matter, Hiroshima University, Higashi-Hiroshima 739-8530, Japan
}%
\author{K. Umeo}%
\affiliation{%
Cryogenics and Instrumental analysis Division, N-BARD, Hiroshima University, Higashi-Hiroshima 739-8526, Japan
}%
\author{S. Kittaka}%
\affiliation{%
Institute for Solid State Physics, University of Tokyo, Kashiwa 277-8581, Japan
}%
\author{T. Sakakibara}%
\affiliation{%
Institute for Solid State Physics, University of Tokyo, Kashiwa 277-8581, Japan
}%
\author{Y. Matsushita}%
\affiliation{%
Research Network and Facility Services Division, National Institute for Materials Science, Sengen, Tsukuba,305-0047 Japan 
}%
\author{T. Takabatake}%
\affiliation{%
Department of Quantum Matter, Graduate School of Advanced Sciences of Matter, Hiroshima University, Higashi-Hiroshima 739-8530, Japan
}%
\affiliation{%
Institute for Advanced Materials Research, Hiroshima University, Higashi-Hiroshima 739-8530, Japan
}%


\date{\today}

\begin{abstract}
Superconducting and antiferroquadrupolar (AFQ) transitions in a Pr-based compound {\prrhzn} have been found to occur simultaneously at {\tc}$=${\tq}$=$0.06 K.
The superconducting transition manifests itself by zero resistance and large diamagnetic susceptibility.
The specific heat exhibits a Schottky anomaly peaking at 14 K and magnetization curves measured at 2 K show anisotropic behaviors. 
The analysis of these data indicates that the crystalline electric field (CEF) ground state of the trivalent Pr ion is the non-Kramers ${\Gamma}_{3}$ doublet with the quadrupolar degrees of freedom.
A sharp peak in the specific heat at 0.06 K has been attributed not to the superconducting transition but to the AFQ transition because the ordering temperature {\tq} decreases in \textbf{\textit B} ${||}$ [100] but increases in \textbf{\textit B}$ {||}$ [110] and \textbf{\textit B} ${||}$ [111]  with increasing $B$ up to 6 T.
This anisotropic behavior of {\tq}($B$) can be well explained by a two-sublattice mean-field calculation, which corroborates the AFQ ordered state below {\tq}.
The entropy release at {\tq} is only 10\% of $R$ln2 expected for the ${\Gamma} _{3}$ doublet, suggesting 
possible interplay between the quadrupolar degrees of freedom and the superconductivity.

\end{abstract}

\maketitle
\section{Introduction} 

Rare-earth intermetallic compounds have attracted much attention because of a variety of phenomena originating from 4$f$ electrons such as heavy-fermion state, superconductivity, Kondo effect, and multipole ordering.
In these systems, the total angular momentum $J$ of the 4$f$ electrons is the good quantum number because of the strong spin-orbital interaction. Therefore, the observable is not the orbital degrees of freedom but quadrupoles which are described as second-order tensors of $J$.
When the dipole magnetic moments are quenched but the quadrupoles remain active under the crystalline electric field (CEF), which is possible in non-Kramers ions at a cubic point group local symmetry, the quadrupoles often play an important role in forming exotic electronic ground states such as antiferroquadrupolar (AFQ) state with a staggered quadrupolar component\cite{Morin82} and a non-Fermi liquid (NFL) state attributed to two-channel (quadrupole) Kondo effect.\cite{Cox87,Cox96,Cox98}
Furthermore, the feasibility of superconductivity mediated by quadrupolar fluctuations in the cubic Pr-based superconductor {\prossb} with {\tc}$=$1.5 K has been pointed out by neutron scattering and NQR measurements.\cite{Bauer02,Kuwahara05,Yogi06}
Another well-known cubic system is PrPb$_{3}$ which undergoes an AFQ transition at {\tq}$=$0.4 K.\cite{Bucher74,Tayama01} 
In this system, the CEF splits the ninefold multiplet of ${J}{=}$4 into four multiplets. 
The CEF ground state of ${\Gamma}_{3}$ doublet carries no magnetic dipole moment but electric quadrupole moments $O_{2}^{0}$=(3$J_{z}^{2}-{\textbf{\textit J}}^{2}$)$/$2  and $O_{2}^{2}$=$\sqrt{3}$($J_{x}^{2}-J_{y}^{2}$)/2.
The former has been found to be the AFQ order parameter by the combined analysis of magnetization and neutron diffraction experiments.\cite{Onimaru04,Onimaru05} The quadrupole moments are aligned with an incommensurate sinusoidally modulated structure even in the ground state. Thereby, the indirect 
Ruderman-Kittel-Kasuya-Yosida-type interaction between the quadrupoles plays the essential role.
On the other hand, substituting La for Pr in {\prpb} destroys
the AFQ order at ${x}{=}$0.03 in Pr$_{1-x}$La$_{x}$Pb$_{3}$.\cite{Kawae06} NFL behavior appearing in the specific heat for ${x}{\ge}$0.95 was attributed to the quadrupole Kondo effect.
The absence of AFQ order between the $\Gamma_3$ doublets in PrInAg$_{2}$ and PrMg$_{3}$ was discussed by taking the quadrupole Kondo effect into consideration.\cite{Yatskar96,Tanida06,Morie09}

We have recently reported that a cubic compound {\prirzn} undergoes an AFQ ordering at {\tq}$=$0.11 K and a superconducting transition at {\tc}$=$0.05 K.\cite{Onimaru10, Onimaru11}
This compound crystallizes in the cubic CeCr$_{2}$Al$_{20}$-type structure with the space group $Fd\bar{3}m$ and ${Z}{=}$8,\cite{Nasch97} where the Pr atom is encapsulated in a highly symmetric Frank-Kasper cage formed by 16 zinc atoms.
The superconductivity below {\tc} was indicated by the sizable diamagnetic signal in the AC magnetic susceptibility.
Our analysis of the magnetic anisotropy in the paramagnetic state and a Schottky peak in the specific heat revealed that the CEF ground state is the non-Kramers ${\Gamma}_{3}$ doublet and the two-fold degeneracy is released by the AFQ ordering at {\tq}$=$0.11 K. 
Thus, this compound is the first example where the superconducting transition occurs in the AFQ ordered state.
The entropy at {\tq} is reduced to 0.2$R$ln2 from $R$ln2 that is expected for two-fold degeneracy of the ${\Gamma}_{3}$ state. This strongly suggests that the quadrupole fluctuations play a role in forming the superconducting Cooper pairs.

Matsushita {\it et al}. have reported that the specific heat has no peak at {\tq} but a broad peak at around 0.4 K for their single-crystalline sample of {\prirzn}\cite{Matsushita11}. It was interpreted as the manifestation of a quadrupole-glass-like ordering.The absence of the sharp peak due to quadrupolar transition, however, might result from the lower quality of their sample than that of ours. 
In fact, the residual resistivity of their sample is 40 times larger than ours.
From the very small effective mass at the Fermi level, they argued very weak hybridization effect of the 4$f$ electrons with the conduction electrons.\cite{Matsushita11} 

It is worthy noting that for our samples of {\prirzn} the specific heat divided by temperature, $C/T$, shows the $-$ln$T$ dependence and the electrical resistivity $\rho(T)$ shows $\sqrt{T}$ dependence in the range between 0.2 K and 0.8 K.\cite{Onimaru11b} These temperature dependences are in accord with the theoretical calculation based on a two-channel Anderson lattice model with a low characteristic temperature.\cite{Tsuruta12} This good accordance suggests that quadrupolar excitations are weakly coupled to the conduction electrons.  We note that the above temperature dependences in $C/T$ and $\rho(T)$ have been reported in an isostructural compound PrV$_2$Al$_{20}$ with the $\Gamma_3$ doublet CEF ground state above {\tq}.\cite{Sakai11} Furthermore, another isostructural compound PrTi$_2$Al$_{20}$ has been found to show a superconducting transition at 0.2 K in the ferroquadrupole ordered state. Thereby, clean limit superconductivity and enhanced effective mass of ${m}^{*}{/}{m}{\sim}$ 16 have been suggested.\cite{Sakai12}
 Therefore, in order to address the issue whether the superconducting pair is mediated by the quadrupole fluctuations or not, further experimental investigations of magnetic and transport properties of the family of PrT$_2$X$_{20}$ (X$=$Al and Zn) are certainly needed.
 
In the present paper, we focus on {\prrhzn}  which is isoelectronic to {\prirzn}. Because similar CEF level scheme is expected for the two systems, the CEF ground state in {\prrhzn} would be
the $\Gamma_3$ doublet. It is intriguing to know whether the AFQ ordering and superconducting transition occur or not in {\prrhzn}. Keeping this in mind, we have grown single-crystalline samples and studied the magnetic and transport properties.
A part of the present work has been published in conference proceedings.\cite{Onimaru11b,Nagasawa11}

\section{Experimental}

Single-crystalline samples of {\prrhzn} were grown by the zinc self-flux method with high purity elements of Pr (4N), Rh (3N) and Zn (6N). The CeCr$_2$Al$_{20}$-type structure was confirmed by the powder x-ray diffraction analysis. By X-ray diffraction analysis for a single-crystalline sample at 293 K with Mo K$\alpha$ radiation using a Bruker SMART APEX CCD area-detector diffractometer, the lattice parameter at room temperature was refined to be 14.2702(3) ${\rm \AA}$, whose value is a little smaller than 14.287(1) ${\rm \AA}$ in the previous report.\cite{Nasch97} It is comparable with that for the isoelectronic {\prirzn}, 14.2729(2) ${\rm \AA}$.\cite{Onimaru10} The chemical compositions of the crystals were determined by averaging over 10 different regions for each crystal with a JEOL JXA-8200 analyzer. The compositions of Pr : Rh : Zn $=$ 1 : 1.98(1) : 20.02(5), where the numbers in the parentheses are the standard deviations. The electron-probe microanalysis revealed the impurity phases of binary alloys RhZn$_{6.4}$ and Pr$_{14}$Zn whose volume fractions are less than 5\% in view of back-scattering electron image. The single-crystalline sample was oriented by the back reflection Laue method using an imaging plate camera, IPXC/B (TRY-SE).

The electrical resistivity $\rho$ was measured by a standard four-probe AC method in a laboratory-built system with a Gifford-McMahon-type refrigerator between 3 and 600 K. The measurements down to 0.045 K were done in a laboratory-built system with a commercial Cambridge Magnetic Refrigeration mFridge mF-ADR50. 
Magnetization was measured using a commercial SQUID magnetometer (Quantum Design MPMS) between 1.9 and 300 K in magnetic fields up to 5 T. 
The AC magnetic susceptibility was measured down to 0.045 K  using a laboratory-built system installed in the mFridge mF-ADR50. The measurements were carried out for a frequency of 1 kHz, with an applied AC field of 0.02 mT.
Single-crystalline samples of {\prrhzn} and the reference superconductor {\lairzn} of the same dimensions 0.5${\times}$1${\times}$4 mm$^3$ were loaded into pickup coils.
The specific heat $C$ was measured by a relaxation method using a commercial calorimeter (Quantum Design PPMS) at temperatures between 0.4 and 300 K. 
The measurements down to 0.047 K were done by a quasi-adiabatic method with a $^3$He-$^4$He dilution refrigerator. 
Thereby, horizontal magnetic field up to 7 T was applied by a split-pair superconducting magnet.
In zero magnetic field measurements, the dilution refrigerator was inserted into a cryo-dewar without a superconducting magnet to avoid residual magnetic fields.

\section{Results and discussion}

\begin{figure}[t]
\includegraphics[scale=0.5]{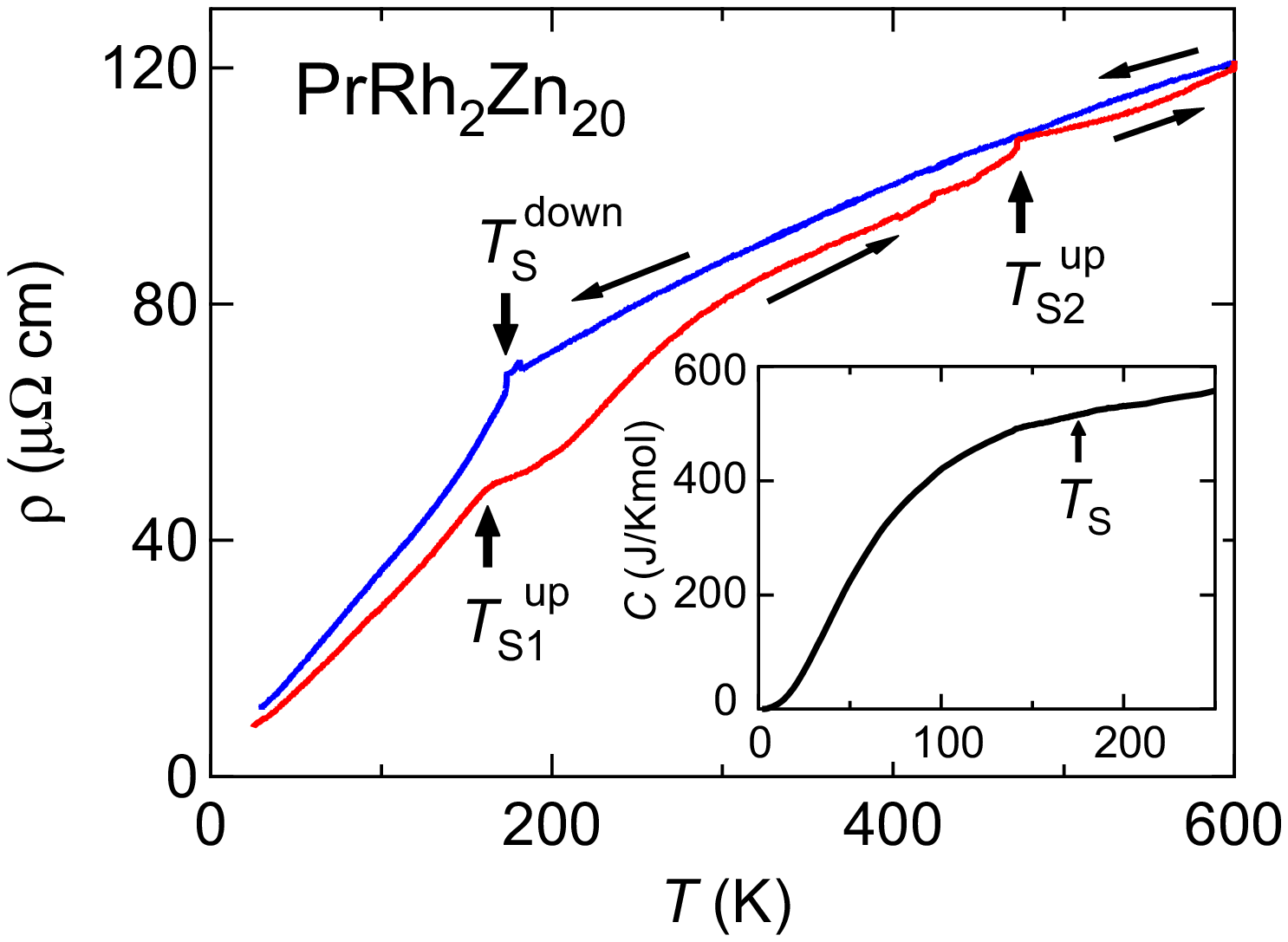}
\caption{(Color online) Temperature dependence of the electrical resistivity of {\prrhzn} in the temperature range between 30 K and 600 K. $T_{\rm S}^{\rm down}$, $T_{\rm S1}^{\rm up}$ and $T_{\rm S2}^{\rm up}$ denote the temperatures of anomalies in the cooling and heating processes, respectively. The inset shows that the specific heat does not exhibit any anomaly at $T_{\rm S}$.}
\label{rho_HC}
\end{figure}

Figure \ref{rho_HC} shows the temperature dependence of the electrical resistivity ${\rho}(T)$ of {\prrhzn} between 30 K and 600 K.
The hysteretic behavior appears in a wide range between 140 and 470 K, which indicates a first-order phase transition. 
However, the specific heat data shown in the inset of Fig. \ref{rho_HC} exhibits no apparent peak at around {\ts}, because the first-order phase transition is hard to be observed by the relaxation method. 
Hysteretic behaviors in ${\rho}(T)$ were also observed in {\laruzn} and {\prruzn} at {\ts}=150 and 138 K, respectively.\cite{Onimaru10} Since superlattice reflections were observed in {\prruzn} below {\ts} by the electron diffraction method, the hysteretic behavior was attributed to a structural transition.\cite{Onimaru10} 
On the other hand, no evidence for structural transition was found in $\rho$($T$) from 530 K to 30 K for {\prirzn}. Therefore, the quadrupolar degrees of freedom remain active in the $\Gamma_3$ doublet.\cite{Onimaru11} It undergoes antiferroquadrupole ordering at {\tq}$=$0.11 K and superconducting transition at {\tc}$=$0.05 K.

\begin{figure}[t]
\includegraphics[scale=0.5]{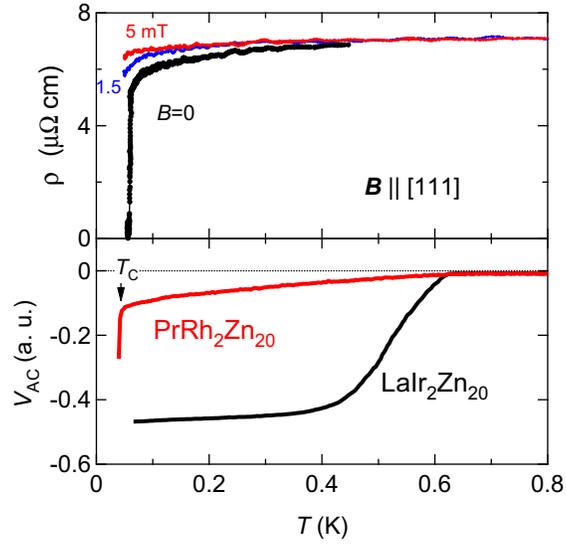}
\caption{(Color online) Temperature dependence of the electrical resistivity $\rho$ in $B$$=$0, and in magnetic field of 1.5 and 5 mT applied along the [111] direction.  The inset shows the AC magnetic susceptibility of {\prrhzn} and the reference compound {\lairzn} with {\tc}$=$0.6 K.}
\label{sc}
\end{figure}

Figure \ref{sc} shows the low-temperature part of ${\rho}(T)$ of {\prrhzn} in various constant magnetic fields $B$$=$0, 1.5, and 5.0 mT applied along the [111] direction.  
In $B$$=$0, $\rho$ drops to zero at {\tc}$=$0.06 K, indicating a superconducting transition.
The bulk nature of the superconductivity was confirmed by detecting a large diamagnetic signal of the Meissner effect below {\tc} in the AC magnetic susceptibility as shown in the inset of Fig. \ref{sc}. The diamagnetic signal below {\tc} for {\prrhzn} is about half that of the reference superconductor {\lairzn} with {\tc}$=$0.6 K, indicating that the superconducting volume fraction in the sample is about 50 \% at the lowest temperature of 0.045 K.
Applying magnetic fields of 1.5 mT and 5.0 mT, the drop of $\rho$ is suppressed and the normal state remains to the lowest temperature 0.05 K.
This means that {\tc} is suppressed below 0.05 K even in the small magnetic field of $B$$=$1.5 mT.
The vanishment of {\tc} in $B$$=$1.5 mT seems to be consistent with the critical magnetic field of 0.5 mT predicted from the BCS theory.
The magnetic field dependence of {\tc} at low fields below 1 mT should be measured in detail to evaluate the critical field. 
Below 0.4 K, the ${\rho}(T)$ increases with increasing magnetic fields, probably because fluctuations of field-induced magnetic moments effectively scatter the conduction electrons.

\begin{figure}[t]
\includegraphics[scale=0.5]{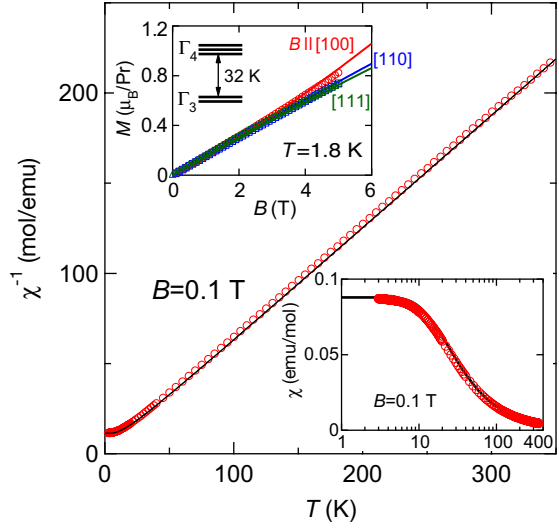}
\caption{(Color online) Temperature dependence of the inverse magnetic susceptibility ${\chi}^{-1}$ of {\prrhzn}. The solid curve is a fit of the data with the CEF parameters of ${W}{=}{-}$1.1 K and ${x}{=}$0.46, and the mean-field parameter ${\lambda}{=}{-}$1.5 mol/emu.
The lower inset shows the magnetic susceptibility ${\chi}(T)$.
The upper inset shows the isothermal magnetization at 1.8 K in the magnetic fields applied along the [100], [110], and [111] axes.}
\label{mag}
\end{figure}

The data of $\chi$($T$) for {\prrhzn} is plotted in the lower inset of Fig. \ref{mag}. On cooling below 10 K, $\chi$($T$) gradually approaches a constant value, indicating a Van-Vleck paramagnetic ground state. Thus, the CEF ground state is nonmagnetic, either $\Gamma_1$ singlet or $\Gamma_3$ doublet. The main panel of Fig. \ref{mag} shows ${\chi}^{-1}$($T$) measured at $B$$=$0.1 T. Above 30 K, ${\chi}^{-1}$ follows the Curie-Weiss law with the effective magnetic moment of 3.55 ${\mu}_{\rm B}$/f.u., which is in agreement with the value of the trivalent Pr free ion. The small value of $-$4.8 K for the paramagnetic Curie temperature ${\theta}_{p}$ indicates that the intersite magnetic interaction between Pr ions is antiferromagnetic but rather weak.

\begin{figure}[t]
\includegraphics[scale=0.5]{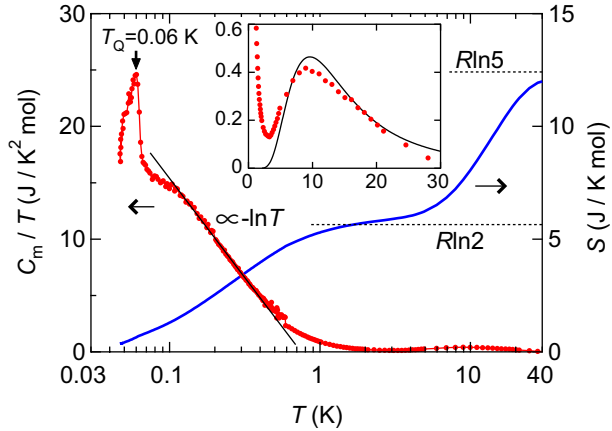}
\caption{(Color online) Temperature dependence of the magnetic specific heat divided by temperature, $C_{\rm m}$/$T$ (left-hand scale) and the entropy (right-hand scale). The inset shows the  $C_{\rm m}$/$T$ on a linear temperature scale.}
\label{HC_B0}
\end{figure}

To judge whether the CEF ground state is the nonmagnetic either $\Gamma_1$ singlet or $\Gamma_3$ doublet, we measured the isothermal magnetization $M(B)$ at $T$$=$1.8 K by applying magnetic fields up to 5 T along the [100], [110], and [111] directions.
As shown in the upper inset of Fig. \ref{mag}, the $M(B)$'s data are almost the same up to 2 T, above which three curves gradually diverge; $M$(\textbf{\textit B} $||$ [100]) $>$ $M$(\textbf{\textit B} $||$ [110]) $>$ $M$(\textbf{\textit B} $||$ [111]). This anisotropic behavior can be well reproduced by the calculation of $M$($B$) using the $\Gamma_3$ $-$ $\Gamma_4$ CEF model, as shown with the solid lines.
The $\Gamma_3$ $-$ $\Gamma_4$ level scheme separated by 32 K is depicted in the inset.
Figure \ref{HC_B0} shows the temperature dependence of the magnetic part of the specific heat divided by temperature, $C_{\rm m}{/}{T}$. For the lattice part of $C$ in {\prrhzn}, we used the data of {\lairzn} because the more relevant compound {\larhzn} does not form. As shown in the inset of Fig. \ref{HC_B0}, a broad peak manifests itself at around 10 K. 
To reproduce this Schottky-like anomaly, we used the $\Gamma_3$ $-$ $\Gamma_4$ model. 
The solid line represents the calculation, where the higher excited levels were not taken into account. 
The good fit corroborates the $\Gamma_3$ doublet ground state. 
The $\Gamma_3$ ground state has been also supported by the inelastic neutron scattering experiments, where the excitations from the $\Gamma_3$ ground state to the excited states were observed.\cite{Iwasa12} 
Furthermore, no additional excitation peak appears below {\ts}, indicating that the cubic point group of the Pr site is conserved even below {\ts}.

\begin{figure}[t]
\includegraphics[scale=0.4]{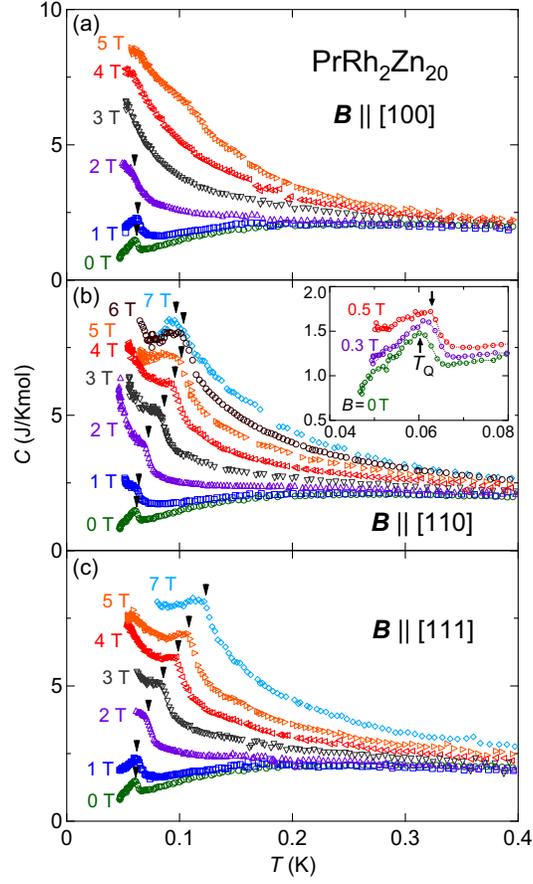}
\caption{(Color online) Temperature dependence of the specific heat in various constant magnetic fields for (a) \textbf{\textit B} $||$ [100], (b) \textbf{\textit B} $||$ [110], and (c) \textbf{\textit B} $||$ [111].}
\label{HC}
\end{figure}

The gradual release of the entropy of the $\Gamma_3$ doublet results in the gradual increase in $C_{\rm m} {/} {T}$ for ${T}{<}$3 K. 
Between 0.6 and 0.1 K, $C_{\rm m}$/$T$ obeys $-$ln$T$ dependence as is shown with the solid line in Fig. \ref{HC_B0}. It is probably the manifestation of the two-channel Kondo effect due to the quadrupolar degrees of freedom as were found in the isostructural {\prirzn} and PrT$_2$Al$_{20}$ (T$=$V and Nb).\cite{Onimaru11b,Sakai11,Higashinaka11}
The slope of the $-$ln$T$ dependence of $C_{\rm m} / T$ in Fig. \ref{HC_B0} gives the characteristic temperature $T_{\rm K}$ as 0.7 K.\cite{Cox98} The entropy release at $T_{\rm K}$ from $R$ln2 is 1.35 J  $/$ K$^2$ mol, which value is moderately consistent with the theoretical value 1.45 J $/$ K$^2$ mol calculated by taking the two-channel Kondo effect into consideration.\cite{Cox98}
On further cooling, a peak appears at 0.06 K, which coincides with the superconducting transition at {\tc}$=$0.06 K. 
To judge whether this peak in $C_{\rm m}{/}{T}$ is due to the superconducting transition or not, we measured the specific heat in magnetic fields. The inset of Fig. \ref{HC} (b) shows the data in $B$$=$0, 0.3, and 0.5 T applied along the [110] direction. The peak at 0.06 K in $B$$=$0 slightly shifts to higher temperatures at $B$$=$0.3 and 0.5 T, indicating the phase transition remains even in $B$$=$0.5 T. 
Note that the resistive transition into the superconducting state is quenched by the very weak field of 1.5 mT as shown in Fig. \ref{sc}.
Assuming a BCS superconductor with electronic specific heat $\gamma$$=$ 10 mJ/K$^2$ mol, the jump in the specific heat, ${\Delta}{C}$, is evaluated as 8 mJ/K mol at {\tc}$=$0.06 K from the relation of ${\Delta}{C}{/}{\gamma}${\tc}$=$1.43.
This value of ${\Delta}{C}$ is smaller than the resolution of our specific heat measurements with the quasi-adiabatic method, say 50 mJ/K mol at 0.06 K.
Therefore, we conclude that the peak of the specific heat at 0.06 K does not originate from the superconducting transition but a phase transition due to the multipolar degrees of freedom in the $\Gamma_3$ doublet ground state. Thus, we mark this transition temperature as {\tq} hereafter.
The magnetic entropy $S$ was evaluated from the temperature dependence of the $C_{\rm m}$/$T$ as shown with the solid curve in Fig. \ref{HC_B0}.
On cooling, the $S(T)$ monotonically decreases from the value of $R$ln2 at around 2 K.
The $S$ at {\tq}$=$0.06 K is only 0.1$R$ln2 which is much smaller than $R$ln2 that is expected for the two- fold degeneracy of the $\Gamma_3$ doublet.

\begin{figure}[t]
\includegraphics[scale=0.55]{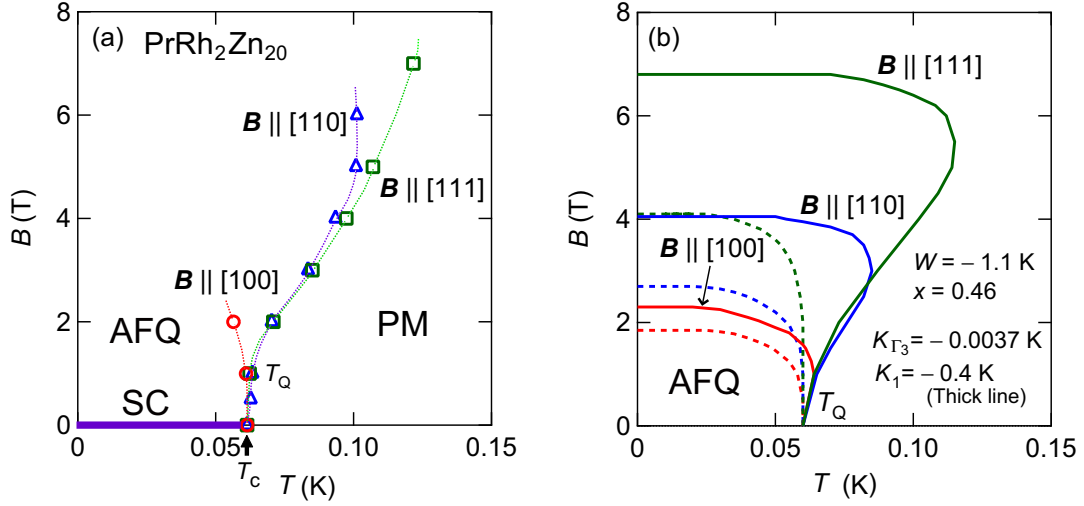}
\caption{(Color online) (a) $B$ $-$ $T$ phase diagrams for \textbf{\textit B} $||$ [100], [110], and [111] determined by the specific heat measurements. {\tq} and {\tc} are the AFQ and superconducting transition temperatures, respectively. AFQ, SC, and PM indicate the antiferroquadrupolar ordering, superconducting, and paramagnetic states, respectively. 
(b) $B$ $-$ $T$ phase diagrams obtained by the two-sublattice mean-field calculation. The CEF parameters are $W$=$-$1.1 K and $x$=0.46, and the quadrupole interaction is $K_{\Gamma_{3}}$=$-$0.0037 K. The solid and dashed curves indicate the phase boundaries calculated with the intersite magnetic interaction $K_1$=$-$0.4 K and without the magnetic interaction $K_{1}{=}$0, respectively. }
\label{Diagram}
\end{figure}

Figure \ref{HC} shows the temperature dependence of the specific heat in magnetic fields applied along (a) [100], (b) [110], and (c) [111].
For \textbf{\textit B} $||$ [100], {\tq} decreases with increasing the magnetic field, and disappears above 3 T.
On the other hand, for \textbf{\textit B} $||$ [110] and [111], {\tq}'s increase with increasing the magnetic fields up to 6 T.
Above 5 T, {\tq} starts decreasing in \textbf{\textit B} $||$ [110], whereas {\tq} still increases in \textbf{\textit B} $||$ [111].
The magnetic field dependences of {\tq} are summarized in the $B$$-$$T$ phase diagrams in Fig. \ref{Diagram} (a), where AFQ, SC, and PM mean the antiferroquadrupolar ordering, superconducting, and paramagnetic states, respectively.
In $B$$=$0, the superconducting transition and AFQ transition occur simultaneously at {\tc}$=${\tq}$=$0.06 K.
Note that {\tq} depends on the magnetic field directions; {\tq} (\textbf{\textit B} $||$ [100]) $<$  {\tq} (\textbf{\textit B} $||$ [110]) $<$  {\tq} (\textbf{\textit B} $||$ [111]).
This relation is similar to those observed in {\prpb} and {\prirzn} with the non-Kramers $\Gamma_3$ doublet ground states.\cite{Tayama01, Onimaru11,Ishii11}

The anisotropic behavior of {\tq}($B$) in {\prrhzn} can be explained by the calculation of the CEF level scheme in the paramagnetic state.
The splitting of the $\Gamma_3$ doublet calculated for \textbf{\textit B} $||$ [100], ${\Delta}_{100}$, is much larger than ${\Delta}_{110}$ in \textbf{\textit B} $||$[110] and ${\Delta}_{111}$ in \textbf{\textit B} $||$[111].
This calculation reproduces the isothermal magnetization at 1.8 K shown in the upper inset of Fig. \ref{mag}, where $M$(\textbf{\textit B} $||$ [100]) exceeds $M$(\textbf{\textit B} $||$ [110]) and $M$(\textbf{\textit B} $||$ [111]) above 2 T. 
If the energy scale of the quadrupole interaction becomes smaller than the field induced splitting of the $\Gamma_3$ doublet, the quadrupole freedom would not order any more. 
Therefore, the order of ${\Delta}_{100}$ $>$ ${\Delta}_{110}$ $>$ ${\Delta}_{111}$ in the magnetic fields corresponds to the order of {\tq}; {\tq} (\textbf{\textit B} $||$ [100]) $<$  {\tq} (\textbf{\textit B} $||$ [110]) $<$  {\tq} (\textbf{\textit B} $||$ [111]).

Furthermore, for the quantitative analysis of the above results, we performed mean-field calculation with a two-sublattice model based on the following isotropic interactions:
\begin{multline}
\mathcal{H}_{\rm A(B)}^{\rm I}=\mathcal{H}_{\rm CEF}-g_{J}{\mu}_{B}\textbf{\textit {JH}}
-(K_{1}\langle\textbf{\textit J}\rangle_{\rm B(A)}
+K_{2}\langle\textbf{\textit J}\rangle_{\rm A(B)})\textbf{\textit J}\\
-K_{\Gamma{3}}[\langle{O_{2}^{0}}\rangle_{\rm B(A)}O_{2}^{0}
+\langle{O_{2}^{2}}\rangle_{\rm B(A)}O_{2}^{2}],\\
\label{mean_field}
\end{multline}
where $g_{J}$ is the Land\'{e} $g$-factor ($=4/5$ for Pr$^{3+}$) and $\mu_{\rm B}$
is the Bohr magneton.
The first term in eq.~(\ref{mean_field}) is the CEF Hamiltonian
for the subspace of the ${J}{=}$4 multiplet, 
which can be written as
\begin{equation}
\mathcal{H}_{\rm CEF}=W[{x}\frac{O_{4}^{0}-5O_{4}^{4}}{60}
+(1-|{x}|)\frac{O_{6}^{0}-21O_{6}^{4}}{1260}],
\label{CEF}
\end{equation}
where the notation by Lea {\it et al.} is used.\cite{Lea62}
Here, we adopt the two parameters ${W}{=}{-}$1.1 K and 
${x}{=}$0.46 which were determined to reproduce both the magnetic susceptibility as shown with the solid curves in Fig. 3 and the inelastic neutron scattering spectra\cite{Iwasa12}. The eq. (\ref{CEF}) leads to the CEF level scheme as $\Gamma_3$ (0) $-$ $\Gamma_4$ (31 K) $-$ $\Gamma_5$ (65 K) $-$ $\Gamma_1$ (73 K).
In eq. (\ref{mean_field}), ${K}_{1}$ and ${K}_{2}$ are the inter- and intra-sublattice 
magnetic interaction coefficients for excited magnetic multiplets, respectively, and
${K}_{\Gamma_{3}}$ is the inter-sublattice interaction coefficient of ${\Gamma}_{3}$-type 
quadrupolar moments.
It should be noted that the quadrupoles $O_2^0$ and $O_2^2$ have the same interaction coefficient in this model. As a result, the
$O_2^0$ and the $O_2^2$ phases are degenerate for $B=0$, i. e. $T_{\rm Q}(O_2^0)=T_{\rm Q}(O_2^2)$.
From the experimental result {\tq}$=$0.06 K for ${B}{=}$0, we obtain ${K}_{\Gamma_{3}}{=}{-}$0.0037 K.
The difference between the measured magnetic susceptibility ${\chi}_{4f}$ at ${T}{>}${\tq} and ${\chi}_{\rm CEF}$ calculated using the CEF parameters shown above leads to the mean-feild parameter ${\lambda}{=}{-}$1.5 mol/emu, which represents the exchange interaction among the Pr magnetic moments, and the relation $K_{1}{+}K_{2}{=}{-}$ 0.4 K. Although these antiferromagnetic interactions are much stronger than the AFQ interaction,
the system does not order antiferromagnetically because the non-magnetic $\Gamma_{3}$ doublet is well separated from the excited magnetic multiplets.

The calculated ${B}{-}{T}$ phase diagrams for \textbf{\textit B} $||$ [100], \textbf{\textit B} $||$ [110], and \textbf{\textit B} $||$ [111] are shown with the thick curves in Fig. \ref{Diagram} (b).
For simplicity, we set $K_{2}{=}$0 in the relation $K_{1}{+}K_{2}{=}{-}$ 0.4 K.
For \textbf{\textit B} $||$ [100], [110], and [111], {\tq} increases with increasing magnetic fields up to ${B}{=}$1, 3, and 5 T, and at ${T}{=}$0 the AFQ ordered phase closes at the critical fields of ${B}{=}$2.3, 4.0, and 6.3 T , respectively.
This calculation of {\tq}($B$) qualitatively reproduces both the anisotropic response of {\tq} and the initial increment of {\tq} with the magnetic fields.
Although the calculated critical fields of the AFQ ordered phase are much lower than the experimental data.
The quantitative disagreement is owing to simple two-sublattice model where only the isotropic magnetic and quadrupole interactions were taken into consideration. 
Quadrupole structures could be more complicated than the two-sublattice model and there exist various kinds of interactions between the higher-order multipoles induced by magnetic field.
In order to deduce the effect of the antiferromagnetic interaction on {\tq}($B$), we calculated {\tq} by setting $K_{1}{=}K_{2}{=}$0.
As is shown  by the dashed curves in Fig. \ref{Diagram} (b),
{\tq}'s do not increase with the magnetic field and the AFQ ordered phase collapses at rather low fields of 1.8, 2.7, and 4.1 T for \textbf{\textit B} $||$ [100], [110], and [111], respectively.
This discrepancy strongly suggests that the AFQ ordered phase is stabilized by the AFM interaction between the field-induced magnetic dipoles.

The important fact of the ${B}{-}{T}$ phase diagram of Fig. \ref{Diagram} (a) is that the superconducting region is inside the AFQ ordered phase as found for the isoelectronic {\prirzn}.\cite{Onimaru11}
As shown in Fig. \ref{HC_B0}, the $S(T)$ at {\tq} is reduced to 0.1$R$ln2 from $R$ln2 that is expected for the two-fold degeneracy of the ${\Gamma}_{3}$ state. 
The reduced entropy at {\tq} in {\prrhzn} from 0.2$R$ln2 in {\prirzn} is attributed to the quadrupole fluctuations.
Although {\tc}$=$0.05 K is lower than {\tq}$=$0.11 K in {\prirzn},\cite{Onimaru11} the two transitions simultaneously occur at {\tc}$=${\tq}$=$0.06 K in {\prrhzn}.
It is interesting to examine if the quadrupole fluctuations would stabilize the superconducting state.
To reveal how the superconductivity couples with the quadrupolar degrees of freedom, a systematic study of the family of {\prtx} (X=Al and Zn) at low temperatures and under high pressures are needed.

It remains as an important issue why the CEF ground state in {\prrhzn} is the $\Gamma_3$ doublet state in spite of the possible structural transition between 140 and 470 K. Because the $\Gamma_3$ doublet is realized only in the cubic point groups, the low temperature phase must belong to another cubic point group.
We recall that the metal-insulator transition in PrRu$_4$P$_{12}$ at $T_{\rm MI}$$=$63 K is accompanied with a structural transition from a body-centered cubic structure in the high-temperature metallic phase to a simple cubic one in the low-temperature insulating phase\cite{Lee01}. In order to determine the structure of the low-temperature phase of {\prrhzn}, x-ray diffraction and electron diffraction measurements on single-crystalline samples are in progress.

\section{Conclusion}

We performed electrical resistivity $\rho$, magnetic susceptibility $\chi$, and specific heat $C$ measurements on {\prrhzn}, which is isoelectronic to {\prirzn} showing antiferroquadrupolar and superconducting transitions at {\tq}$=$0.11 K and {\tc}$=$0.05 K, respectively. 
In {\prrhzn},  we found a superconducting transition at {\tc}$=$0.06 K, below which the bulk nature of the superconductivity was confirmed by the large diamagnetic signal due to the Meissner effect.
The analysis of both the Schottky anomaly in specific heat and anisotropic magnetization curves for ${B}{>}$2 T at ${T}{=}$1.8 K indicates that the CEF ground state of the trivalent Pr ion is the non-Kramers ${\Gamma}_{3}$ doublet with the quadrupolar degrees of freedom.
The sharp peak in the specific heat at 0.06 K is attributed to the antiferroquadrupolar ordering. This temperature {\tq} shows anisotropic response to magnetic field $B$; {\tq} decreases in \textbf{\textit B} ${||}$ [100], whereas {\tq} increases in \textbf{\textit B} ${||}$ [110] and \textbf{\textit B} ${||}$ [111] with increasing $B$ up to 6 T.
This anisotropic behavior of {\tq}($B$) can be well explained by the mean-field calculation with a two-sublattice model,
assuming the isotropic inter-sublattice interaction between the ${\Gamma}_{3}$-type quadrupolar moments and the magnetic interaction between the field-induced magnetic dipole moments. 
These experimental and calculated results corroborate the AFQ ordering below {\tq}.
The superconducting transition and AFQ order occur simultaneously at {\tc}$=${\tq}$=$0.06 K in ${B}{=}$0. 
By applying magnetic fields, however, the superconducting state is easily quenched at several mT but the antiferroquadrupolar phase is stabilized. 
To answer the issue whether the superconducting pair is mediated by the quadrupole fluctuations or not, further experimental and theoretical works are needed.

\section*{ACKNOWLEDGMENTS}

The authors would like to thank I. Ishii, T. Suzuki, K. Iwasa, Y. Saiga, F. Iga, A. Tsuruta, and K. Miyake for helpful discussions.
We also thank Y. Shibata for the electron-probe microanalysis performed at N-BARD, Hiroshima University. 
The magnetization measurements with MPMS and specific heat measurements with PPMS were carried out at N-BARD, Hiroshima University. 
This work was carried out by the joint research in the Institute for Solid State Physics, the University of Tokyo. 
This work was financially supported by Grants-in-Aid from MEXT of Japan, Nos. 20102004, 21102516, 23102718, and 23740275 and by The Mazda Foundation Research Grant, Japan.


\newpage

\end{document}